The evolution of the electron number density in the coma of comet 67P
at the location of Rosetta from 2015 November through 2016 March

E. Vigren[1], N. J. T. Edberg[1], A. I. Eriksson[1], M. Galand[2], P. Henri[3], F. L. Johansson[1],
E. Odelstad[4], M. Rubin[5] and X. Vallières[3]

1) Swedish Institute of Space Physics, Uppsala, Sweden. E-mail: erik.vigren@irfu.se
2) Department of Physics, Imperial College London, London, UK
3) Laboratoire de Physique et Chimie de l'Environnement et de l'Espace, Orleans, France
4) Department of Space and Plasma Physics, Royal Institute of Technology, Stockholm, Sweden
5) Physikalisches Institut, Universität Bern, Switzerland

ABSTRACT

A comet ionospheric model assuming the plasma to move radially outward with the same bulk speed as the neutral gas and not being subject to severe reduction through dissociative recombination has previously been tested in a series of case studies associated with the Rosetta mission at comet 67P/Churyumov-Gerasimenko. It has been found that at low activity and within several tens of km from the nucleus such models (which originally were developed for such conditions) generally work well in reproducing observed electron number densities, in particular when plasma production through both photoionization and electron-impact ionization is taken into account. Near perihelion, case studies have, on the contrary, showed that applying similar assumptions overestimates the observed electron number densities at the location of Rosetta. Here we compare ROSINA/COPS driven model results with RPC/MIP derived electron number densities for an extended time period (2015 November through 2016 March) during the post-perihelion phase with southern summer/spring. We observe a gradual transition from a state when the model grossly overestimates (by more than a factor of 10) the observations to being in reasonable agreement during 2016 March.

1. INTRODUCTION

The Rosetta mission to comet 67P/Churyumov-Gerasimenko (henceforth 67P) gave the opportunity to study the evolution of a cometary coma during a significant part of its orbit around the Sun (currently characterized by a period of 6.44 years and perihelion/aphelion near ~1.25/5.68 AU). Rosetta approached 67P in August 2014 at a heliocentric distance of $d$~3.6 AU. It followed the comet, typically at cometocentric distances, $r$, of tens to hundreds of km, to perihelion in 2015 August and then post-perihelion until the end of the extended mission phase in the end of 2016 September ($d$~3.8 AU).

With changing heliocentric distance, the activity and outgassing pattern of the comet changed dramatically, as reported by, e.g., Hansen et al. (2016) and Läuter et al. (2017). The outgassing was mostly dominated by $H_2O$, in a global sense in particular during southern summer (between the equinoxes in 2015 May 10 and 2016 March 21) with "post-perihelion exceptions" for late parts of 2016 February and 2016 March when the $CO_2$ outgassing was compareble with the $H_2O$ outgassing at southern latitudes (see Fig. 4 in Gasc et al., 2017).

The coma is subject to solar extreme ultraviolet radiation, electron impact, and solar-wind interactions (e.g., Galand et al., 2016, Simon Wedlund 2019) giving rise to a partially ionized coma, the ionization degree of which to first approximation is expected to increase with increasing cometocentric distance and with decreasing heliocentric distance. The cometary ionosphere of 67P has been modeled through MHD- (e.g., Rubin et al., 2014), hybrid- (e.g., Koenders et al., 2015), and Particle-In-Cell (Deca et al., 2017) simulations, while thus far, quantitative comparisons of observed and model-predicted electron number densities are lacking, in part due to limited spatial resolution in the simulations. Analytical models based on the critical simplifying assumption that ions flow radially outward at the same bulk speed as the expanding neutral gas (~0.4-1 km s$^{-1}$, see e.g., Hansen et al., 2016) have been applied in case studies by e.g., Vigren et al. (2016, 2017), Galand et al. (2016) and Heritier et al. (2017, 2018). These focused either on relatively low activity levels (pre-perihelion; Vigren et al. 2016; Galand et al., 2016, post-perihelion; Heritier et al., 2017, 2018), or near perihelion (e.g., Vigren et al., 2017 though the focus of that study was on effective ion speeds). We may also refer to Henri et al. (2017) and Odelstad et al. (2018) for studies of plasma properties within the diamagnetic cavity of 67P. The former study shows that the electron number density is more stable within the cavity than outside, the latter revealing that the flow speed of the ions typically was at the level of several km s$^{-1}$ with a dominating radial component.

The picture that has emerged from these case-studies is that the simplified model (in which the bulk ion velocity, $u_i$, equals the bulk neutral velocity, $u_n$ and dissociative recombination is negligible) works well at low activity (for which it was developed), at least within several tens of km (up to ~70 km, Heritier et al., 2018) from the nucleus. Near perihelion (and when the spacecraft was located beyond a cometocentric distance of 100 km), making similar assumptions, yields an overestimation of the electron number density, compared with the observed ones typically by a factor of 2-5. As for the low activity cases an interesting finding reported by Galand et al. (2016) and Heritier et al. (2018) is that at the location of Rosetta electron-impact ionization often dominates over photoionization as the main source of local ionization and that its incorporation often is required in order for modeled electron number densities to reach the level of the observations. The high flux of energetic electrons has been suggested to result from wave-particle interactions (Broiles et al., 2016) or solar wind electrons that have been accelerated by an ambipolar electric field (Madanian et al., 2016, Deca et al., 2017). Another interesting finding concerns the electron density profile observed during the final descent of Rosetta towards the nucleus surface marking the end of the mission (on 2016 September 30). The density profile was found by Heritier et al. (2017) to match model results (model refined by taken into account a non-constant radial expansion velocity) very well and to include the "predicted" (e.g., Mendis et al., 1981; Vigren & Galand, 2013) plasma peak about a cometary radius above the surface.

The assumptions made at low activity seem to break near perihelion and within the diamagnetic cavity, and this may be caused by:

i) Ion acceleration by the ambipolar electric field (e.g., Vigren & Eriksson, 2017). Within the cavity such a process is not complicated by the presence of a magnetic field.

ii) The neglect of dissociative recombination in the model (e.g., Heritier et al., 2018; Beth et al., 2019).

iii) The neglect of EUV attenuation, particularly by nanograins, which from the work of Johansson et al. (2017) on RPC/Langmuir probe photoelectric emission may be pronounced (up to ~50% reduction).

iv) A combination of the above (e.g., Henri et al., 2017; Heritier et al., 2018).

We may add the possibility of electron depletion caused by nanograin charging as observed in the plume of Enceladus (Morooka et al., 2011) and as investigated for 67P in the parameter study by Vigren et al. (2015b). Except for the few observations of direct Rosetta detections of energetic nanograins reported by Burch et al. (2015) and discussed further by Gombosi et al. (2015), the current evidence for their presence in the 67P environment rather points to their prevalence only at large distances from the nucleus (Boehnhardt et al, 2016; Johansson et al., 2017). Nevertheless, the usually strongly negative spacecraft potential (Odelstad et al., 2015; 2017) could possibly prohibit negatively charged nanongrains at low relative speed to reach the spacecraft, so there remains a slight possibility that their numbers could be underestimated. Finally, outside the diamagnetic cavity, assuming a radial flow may be limited; i.e., the source region of the local (and instantaneous) plasma population may not be pre-dominantly radially inwards. Indeed, both the MHD simulations of Rubin et al. (2012) and the hybrid simulations of Koenders et al. (2015) predict the plasma flow to be bent tailward outside the diamagnetic cavity with bulk velocities (dominated by an anti-sunward component) exceeding 5 km s$^{-1}$ already within several tens of km from the cavity boundary (see e.g., Fig. 3 in Rubin et al., 2012, and Figs. 5 and 8 in Koenders et al., 2015).

While one may ask why $u_i \neq u_n$ near perihelion one may also ask why the simplified model works at low activity. The explanation proposed by e.g., Galand et al. (2016) and Vigren et al. (2016) was that the measurements were conducted inward of the theorized ion-neutral decoupling distance, $r_{in}$. Vigren et al. (2016) leaned on Eq. 10 in Gombosi (2015) which is equivalent with

$$r_{in} = k_{in}r^2 n_\text{n}/u_\text{n} \qquad (1)$$

where $k_{in}$ is the ion-neutral collision rate coefficient [on the order of $10^{-9}$ cm$^3$s$^{-1}$, see e.g., Cravens & Körozmezey, 1986], $r$ is the cometocentric distance, $n_\text{n}$ is the neutral number density at that distance, and $u_n$ is the expansion velocity. A critical weakness in this formulation is that it implicitly assumes $u_\text{i}=u_\text{n}$, a point discussed in more depth in Vigren & Eriksson (2019) who suggest also an alternative definition of $r_\text{in}$ noticing that it is strongly dependent on the electric potential profile (unfortunately poorly constrained). It should be emphasized that Gombosi (2015) stressed that solar-wind interactions had not been taken into consideration when deriving his Eq. 10.

In the present study we aim at providing further clues towards understanding the ionization balance of cometary comae by inspection of electron number densities measured during nearly five months of the Rosetta mission covering a period not much dealt with in previous case studies. We look at the post-perihelion phase from 2015 November through 2016 March 21 witnessing a transition from gross failure of the $u_\text{i}=u_\text{n}$ assumption (and/or additional ones made in the simplified model) to reasonable reproduction of observed electron number densities. The method is described in Section 2, results are presented and discussed in Section 3 and concluding remarks are given in Section 4.

2. METHOD

We refer to Balsiger et al. (2007) and Trotignon et al. (2007) for descriptions of the Rosetta Orbiter Spectrometer for Ion and Neutral Analysis/Comet Pressure sensor (ROSINA/COPS) and the Rosetta Plasma Consortium/Mutual Impedance Probe (RPC/MIP), respectively. Measurements by COPS and MIP yield, respectively, the neutral number densities, $n_\text{n}$, and the electron number densities, $n_\text{e}$, used in this study. We may refer to Galand et al. (2016) and Odelstad et al. (2018) for brief descriptions of the instruments and operational modes and to Carr et al. (2007) for a concise description of all of the instruments within the RPC. During the investigated time period MIP operated pre-dominantly in Short Debye Length (SDL) mode (typically with a cadency of 10 s, though during times in burst mode with a cadency of 3 s) with intermittent periods of Long Debye Length (LDL) measurements. In the LDL mode MIP cannot detect plasma densities higher than ~350 cm$^{-3}$ as limited by the upper frequency range of 168 kHz and we discard these measurements for this study.

The modeled electron number densities are calculated from the COPS neutral number densities (typical cadency of 60 s), $n_\text{n}$, and the cometocentric- and heliocentric distance of Rosetta ($r$ and $d$) via Eq. 2 in Vigren et al. (2016). In this equation, with $d$ inserted in AU and $r$ in km, the electron number density relates to the neutral number density as:

$$n_{e,model} = n_n \frac{r}{d^{3/2}} 10^{-6} \qquad (2)$$

The relation is derived from the probability that an H$_2$O molecule undergoes photoionization prior to reaching the spacecraft [travel time (~$r/u_\text{n}$, with $u_\text{n}$ the expansion velocity) multiplied by ionization frequency]. The $d^{-3/2}$ dependence stem from the $d^{-2}$ dependence of the photoionization frequency and an assumed $d^{-1/2}$ - dependence of the neutral expansion velocity (see Eq. 3. in Cochran & Schleicher, 1993). The factor $10^{-6}$ follows by reducing a pre-factor in Eq. 3 of Cochran & Schleicher (1993) from 0.85 to 0.70 (justified in Vigren et al., 2016) and assuming a constant H$_2$O photoionization frequency at 1 AU of $7\times10^{-7}$ s$^{-1}$ (e.g., Vigren et al., 2015a). Such a crude treatment of the photoionization frequency gives values within 20% of those calculated (from TIMED/SEE L3 V12 data) by Heritier et al. (2018) over the considered time interval (see their Fig. 3). The modified version of Eq. 3 in Cochran & Schleicher (1993) gives neutral velocities in the range ~0.57 km s$^{-1}$ (at ~1.5 AU) to 0.45 km s$^{-1}$ (at ~2.5 AU) over the investigated time period, which is roughly consistent with some pre-perihelion expansion velocities inferred from MIRO measurements (Biver et al., 2015) and ~10-30% lower than the values obtained using Eq. 7 of Hansen et al. (2016). Judging from Fig. 16 of Heritier et al. (2018), with possible exception for 2016 March, we deem it unlikely that electron-impact ionization contributed significantly to the total ionization rate over the investigated time period (inclusion would lead to increased modelled electron densities and therefore to increased modeled-to-observed electron number density ratios). For completeness, we should say that over times (duration of hours to days) the coma was perturbed by transient events like impacts of co-rotating interaction regions, coronal mass ejections, solar flares, and outbursts (Edberg et al., 2016a; 2016b; 2019, Hajra et al., 2017; 2018). In this work we make no attempt to correct for their effect on the ionization balance and we note also that major events dealt with in the above cited studies mainly concern times outside the time interval focused on in this study.

As the plasma density is typically much more variable than the neutral density (e.g., Edberg et al., 2015; Henri et al., 2017) comparisons of modeled plasma densities with observations are not particularly useful over short time-scales (as in seconds). We therefore make use of "median-extraction" of 100 consecutive MIP data points, implying "averaging" over typically ~5-15 minutes and setting the "time stamp" to the average starting times of the individual measurements. The position of Rosetta relative to the rotating nucleus does not change significantly over such time frames, and neither does the neutral density measured by COPS meaning that in a statistical sense it makes sense to use the MIP "average time stamps" to find the associated position of Rosetta (e.g., cometocentric distance and latitude) and the neutral number density via time interpolation of ephemerides and COPS data, respectively. This allows, via Eq. 2, for a simple computing of the ratio of modeled over observed electron number densities.

## 3. RESULTS AND DISCUSSION

Figure 1a gives information on the heliocentric distance, the cometocentric distance, and the latitude of Rosetta during the investigated time period. Note that the selected time period coincides with "southern summer" with generally higher outgassing over southern latitudes. Figure 1b shows the neutral number densities measured by COPS (grey), modelled (black) and observed (red) electron number densities. The displayed MIP data are median values from 100 consecutive measurements in SDL mode. The variation in the modelled electron number densities reflects variations in $r$, $d$, and $n_n$; the latter peaks at southern latitudes and dips at northern. Pronounced spikes in the modelled electron densities may generally be ignored as they are not related to actual abrupt changes in the cometary outgassing but rather to spacecraft maneuvers. Figure 1c shows modeled over observed electron densities (black points) with three horizontal blue lines as guidance for ratios of 1, 3, and 5, respectively.

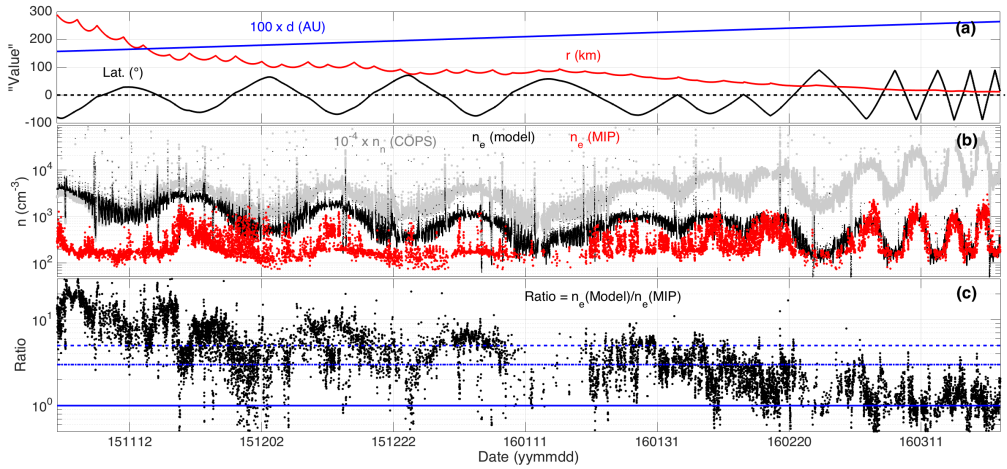

**Figure 1:** (a) Cometocentric distance (red), heliocentric distance (blue, multiplied by 100), and latitude of Rosetta versus time from 2015 November 1 to 2016 March 23 (a time period with southern summer/spring). (b) Neutral number density measured by COPS is shown in gray (multiplied by $10^{-4}$ to fit in the same plot as the other displayed data). Modeled (Eq. 2) and observed (MIP) electron densities are shown in black and red, respectively. The displayed MIP data is the median of 100 consecutive measurements in SDL mode. (c) Modeled to observed electron number density ratios. Here we use the MIP median values shown by red points in Fig. 1b and for the modeled values we interpolate from the black points in Fig. 1b. A few horizontal lines are displayed for guidance of ratios equal to 1, 3, and 5.

The picture that emerges is that the closer to perihelion and further from the nucleus the worse is the agreement between modeled and observed electron number densities. From 2016 mid February, when at a heliocentric distance $d > 2.4$ AU and a cometocentric distance $r < 40$ km (until 2016 March 22), the modelled electron number densities match reasonably the average MIP densities and there is no question that there is a strong correlation between the neutral number density and the average electron number density as also have been emphasized from previous case studies (e.g., Heritier et al., 2018). The interval from 2016 February 20 through 2016 March 21 contains 2,529 SDL extracted median values associated modeled-to-observed plasma density ratios with a median of 1.08, a mean of 1.17 and a standard deviation of the mean of 0.54.

Attributing good model-observation agreement as indicative of strong ion-neutral collisional coupling is questionable. Such a relation would imply that the further inward of the theorized ion-neutral decoupling distance the plasma is sampled, the closer to 1 would the modeled-to-observed electron number density ratio be, at least if neglecting plasma loss through dissociative recombination. From the work of Heritier et al. (2018) it should be fine to neglect dissociative recombination for the considered time interval, with possible exception for parts of 2015 November where the relative error on modeled electron densities (of not including the process) may reach 50% (see their Fig. 6). Figure 2 shows $n_{e,model}/n_{e,MIP}$ ratios (based on averaged values) versus $r/r_{in}$ with $r_{in}$ calculated via Eq. 1 with $k_{in}$ set to $1.5 \times 10^{-9}$ cm$^3$s$^{-1}$ and with $u_n$ set to $u_n = 0.7 \times (d/1 \text{ AU})^{-1/2}$ km s$^{-1}$ (see Section 1). While this form of $r_{in}$ is questionable (Vigren & Eriksson, 2019) the point we are trying to make is not affected by down-scaling the $r_{in}$ value by a multiplicative positive factor smaller than 1. It is seen in Fig. 2 that the $n_{e,model}/n_{e,MIP}$ ratio, if anything, shows an increasing trend with decreasing $r/r_{in}$ ratios. In other words, Fig. 2 gives (surprisingly) no support for the $u_i = u_n$ assumption to hold better within, or even well within the ion-neutral decoupling distance. While the discrepancy between modelled and observed electron densities partly can be due to break-down of other assumptions than $u_i = u_n$ we encourage the search for other plausible explanations (in addition to collisional coupling) as to why the bulk $u_i$ and $u_n$ are so similar at low activity and close to the nucleus. Figure 3 is a density plot of the scattered data in Fig. 2 verifying that the bulk of the data is not concentrated in a small region of the parameter space.

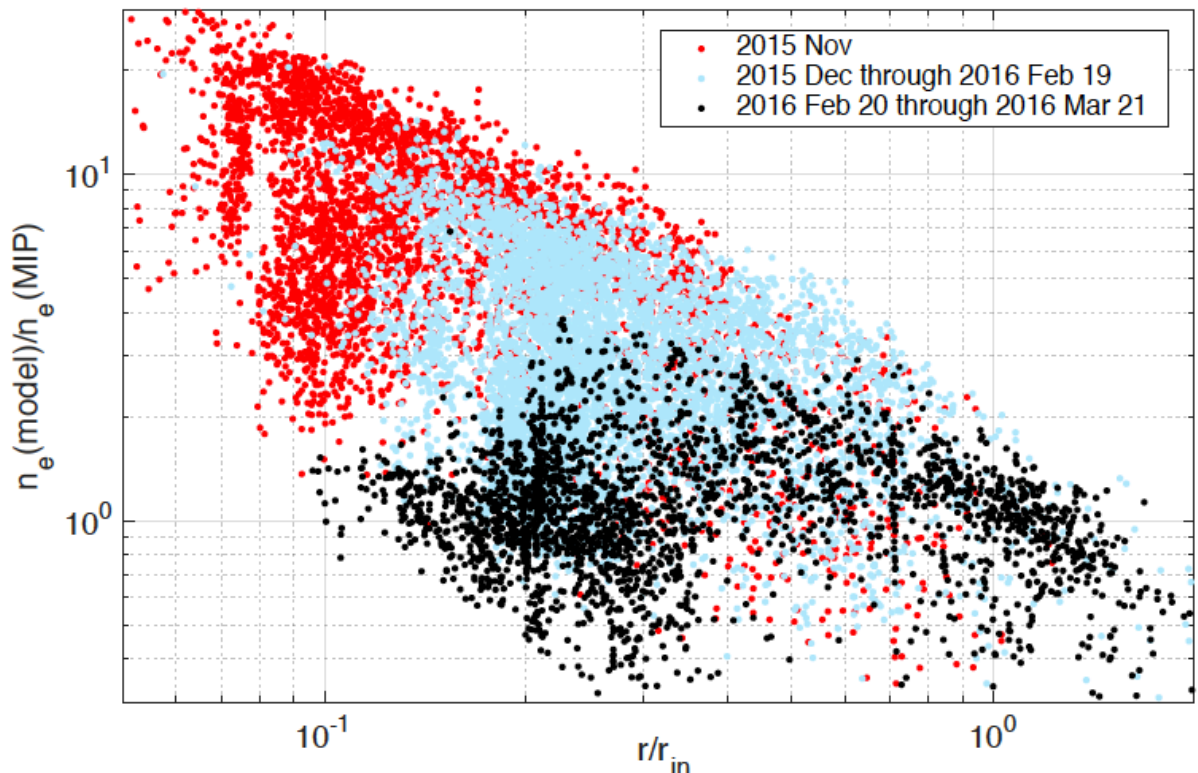

**Figure 2:** Modelled to observed electron number density ratio versus the ratio of the spacecraft-comet distance over the theorized ion-neutral decoupling distance. The MIP data are averages from SDL-mode measurements. The data is divided into three different time periods as indicated in the legend. The figure gives no support for the idea that within the ion-neutral decoupling distance the ion bulk flow velocity may be assumed similar to the neutral flow velocity, as the pronounced deviations in the first and second interval (red and light blue, respectively) hardly can be attributed to the neglect of dissociative recombination in the model.

The previously addressed short time scale variability in $n_e$ (e.g. Edberg et al., 2015 who focused on an early phase of the active mission) and the typically pronounced energy spread of low energy ions measured by RPC/Ion Composition Analyzer (Stenberg Wieser et al., 2017) also speak against a strong ion-neutral coupling at the location of Rosetta. From relative abundances of ion species as measured by the ROSINA/Double Focusing Mass Spectrometer (ROSINA/DFMS) one may make arguments for and against ion-neutral coupling. On the one hand, close to perihelion, detection of $NH_4^+$, at times in comparable (or higher) abundances as (than) $H_2O^+$ and $H_3O^+$ favor an efficient ion-neutral chemistry as $NH_4^+$ is expected to form pre-dominantly via a series of two ion-neutral reactions (Beth et al., 2016). On the other hand, at lower activity and closer to the nucleus, the variable and often low $H_3O^+/H_2O^+$ number density ratios measured by ROSINA/DFMS (Fuselier et al., 2015) does not really line up with the ions being collisionally coupled to the neutrals. To clarify this; chemical models (e.g., Vigren & Galand, 2013; Fuselier et al., 2015; Heritier et al., 2017; Vigren 2018; Beth et al., 2019) that run with the assumption that the ions are cold and moves radially outward with the neutral gas, predicts $H_3O^+$ to most often largely dominate over $H_2O^+$ at the location of Rosetta and this has no strong support in the ROSINA/DFMS ion measurements at low to moderate activity. It should be mentioned that further investigations are needed to come to an understanding of the observed highly variable $H_3O^+/H_2O^+$ ratio (in particular the highly variable $H_2O^+$, Beth et al., 2016) to confirm whether or not it actually is due to a lack of collisional coupling.

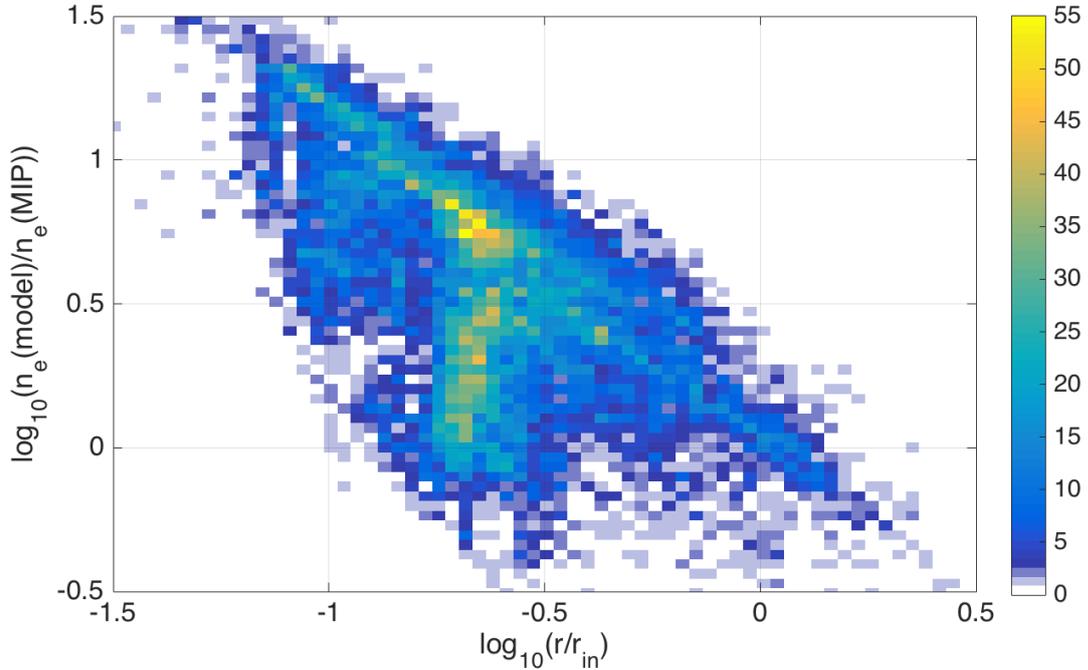

**Figure 3:** Density plot of modeled to observed electron number density ratios versus $r/r_{in}$ (based on the scattered data in Fig. 2). The color represents the number of data points within each bin.

## 4. CONCLUDING REMARKS

We have combined RPC/MIP and ROSINA/COPS data to look further at the evolution of the electron number density (in the coma of 67P at the location of Rosetta) and how it relates to variations in the neutral number density. Our study focuses on an interval not covered extensively in previous case studies and we are witnessing a transition from a scenario wherein a simplified model (developed for low activity) grossly overestimates the observed electron number densities (possibly for reasons discussed in Section 1) to a scenario during lower outgassing where the same type of model reproduces the observations rather well. We find, if anything, an anti-correlation between the modeled-to-observed ratios and the $r/r_{in}$ ratios (where $r$ is the cometocentric distance of Rosetta and $r_{in}$ is the theorized ion-neutral decoupling distance).

The model-observation discrepancy, particularly at high activity, can have multiple causes (e.g., efficient ion-electron dissociative recombination or presence of nanograins causing solar EUV attenuation and electron depletion via grain charging, c.f., Beth et al., 2019; Henri et al., 2017; Heritier et al., 2018; Johansson et al., 2018; Vigren et al., 2015b) and should not only be attributed to a breakdown of the assumption of collisionally coupled ions (with accelerated ions and/or a non radial plasma flow as a result). If ion-neutral decoupling is a prime cause for the discrepancy at moderate activity, then, in light of the above-described anti-correlation, one should seek alternative theories, in addition to ion-neutral coupling, in order to explain why the bulk ion- and neutral velocities are so similar at *low* activity and within several tens of km of the nucleus as shown previously by Galand et al. (2016), Vigren et al. (2016), and Heritier et al. (2018). Based on observations by other Rosetta instruments, in particular RPC/ICA (Stenberg Wieser et al., 2017), we envisage an explanation where the "$u_i$=$u_n$ component" only concerns the bulk radial flow velocity of the ions while the actual mean speed of the ions can be significantly higher than the mean speed of the neutrals.

ACKNOWLEDGEMENTS

Rosetta is a European Space Agency (ESA) mission with contributions from its member states and the National Aeronautics and Space Administration (NASA). Work at the Swedish Institute of Space Physics is supported by the Swedish National Space Board (SNSB) (contract numbers 109/02, 114/13, 135/13, 166/14) and by the Swedish Research Council (contract numbers 621-2013-4191 and 621-2014-5526). Work at Imperial College London is supported by STFC of UK under grant ST/N000692/1. Work at the University of Bern on ROSINA/COPS is funded by the State of Bern, the Swiss National Science Foundation, and the European Space Agency PRODEX Program. Work at LPC2E/CNRS is supported by CNES and ANR under the financial agreement ANR-15-CE31-0009-01.

REFERENCES

Balsiger, H., Altwegg, K., Bochsler, P., et al. 2007, SSRv, 128, 745

Beth, A., Altwegg, K., Balsiger, H, et al. 2016, MNRAS, 462, S562

Beth, A., Galand, M., & Heritier, K. L. 2019, A&A, Accepted, doi: https://doi.org/10.1051/0004-6361/201833517

Biver, N., Hofstadter, M., Gulkis, S., et al. 2015, A&A, 583, A3

Boehnhardt, H., Riffeser, A., Kluge, M., et al. 2016, MNRAS, 462, S376

Broiles, T., Burch, J. L., Chae, K., et al. 2016, MNRAS, 462, S312

Burch, J. L., Gombosi, T. I., Clark, G., Mokashi, P., & Goldstein, R. 2015, GRL, 42, 6575

Carr, C., Cupido, E., Lee, C. G. Y., et al. 2007, SSRv, 128, 629

Cochran, A. L., & Schleicher, D. G. 1993, Icar, 105, 235

Cravens, T. E., & Körosmezey, A. 1986, P&SS, 34, 961

Deca, J., Divin, A., Henri, P., et al. 2017, PhRvL, 118, 205101

Edberg, N. J. T., Eriksson, A. I., Odelstad, E., et al. 2015, GRL, 42, 4263

Edberg, N. J. T., Alho, M., André, M., et al. 2016a, MNRAS, 462, S45

Edberg, N. J. T., Eriksson, A. I., Odelstad, E., et al. 2016b, JGR Space Phys., 121, 949

Edberg, N. J. T., Johansson, F. L., Eriksson, A. I., et al. 2019, A&A (forthcoming), https://doi.org/10.1051/0004-6361/201834834

Fuselier, S., Altwegg, K., Balsiger, H., et al. 2015, A&A, 583, A2

Fuselier, S. A., Altwegg, K., Balsiger, H., et al. 2016, MNRAS, 462, S67

Galand, M., Heritier, K. L., Odelstad, E., et al. 2016, MNRAS, 462, S331

Gasc, S., Altwegg, K., Balsiger, H., et al. 2017, MNRAS, 469, S108

Gombosi, T. I. 2015. Physics of Cometary Magnetospheres, in Magnetotails in the Solar System, edited by A. Kelling, C. M. Jackman and P. A. Delamere, AGU Monograph, 207, pp 169-188

Gombosi, T. I., Burch, J. L., & Horányi, M. 2015, A&A, A23

Hajra, R., Henri, P., Myllis, M., et al. 2018, MNRAS, 480, 4544

Hajra, R., Henri, P., Valières, X., et al. 2017, A&A, 607, A34

Hansen, K. C., Altwegg, K., Berthelier, J.-J., et al. 2016, MNRAS, 462, S491

Henri, P., Vallières, X., Hajra, R., et al. 2017, MNRAS, 469, S372

Heritier, K. L., Galand, M., Henri, P., et al. 2018, A&A, 618, A77

Heritier, K. L., Henri, P., Vallières, X., et al. 2017, MNRAS, 469, S118

Johansson, F. L., Odelstad, E., Paulsson, J. J. P., et al. 2017, MNRAS, 469, S626

Koenders, C., Glassmeier, K.-H., Richter, I., et al. 2015, P&SS, 105, 101

Läuter, M., Kramer, T., Rubin, M., & Altwegg, K. 2019, MNRAS, 483, 852

Madanian, H., Cravens, T. E., Rahmati, A., et al. 2016, JGR Space Phys., 121, 5815

Mendis, D. A., Hill, J. R, Houpis, H. L. F., & Whipple, E. C. 1981, ApJ, 249, 787

Morooka, M. W, Wahlund, J.-E., Eriksson, A. I., et al. 2011, JGR, 116, A12221

Odelstad, E., Eriksson, A. I., Edberg, N. J. T., et al. 2015, GRL, 42, 10126

Odelstad, E., Eriksson, A. I., Johansson, F. L., et al. 2018, JGR Space Phys., 123, 5870

Odelstad, E., Stenberg Wieser, G., Wieser, M., et al. 2017, MNRAS, 469, S568

Rubin, M., Hansen, K. C., Combi, M. R., et al. 2012, JGR Space Phys., 117, A06227

Rubin, M., Koenders, C., Altwegg, K., et al. 2014, Icar, 242, 38

Simon Wedlund, C., Behar, E., Nilsson, H., et al. 2019, A&A, (forthcoming), https://doi.org/10.1051/0004-6361/201834881

Stenberg Wieser, G., Odelstad, E., Wieser, M., et al. 2017, MNRAS, 469, S522

Trotignon, J. G., Michau, J. L., Laboutte, D., et al. 2007, SSRv, 128, 713

Vigren, E. 2018, A&A, 616, A59

Vigren, E., Altwegg, K., Edberg, N. J. T., et al. 2016, AJ, 152, 59

Vigren, E., André, M., Edberg, N. J. T., et al. 2017, MNRAS, 469, S142

Vigren, E & Eriksson, A. I. 2017, AJ, 153, 150

Vigren, E & Eriksson, A. I. 2019, MNRAS, 482, 1937

Vigren, E., & Galand, M. 2013, ApJ, 772, 33

Vigren, E., Galand, M., Eriksson, A. I., et al. 2015a, ApJ, 812, 54

Vigren, E., Galand, M., Eriksson, A. I., et al. 2015b, ApJ, 798, 130